\documentclass[twocolumn,aps,pre,showpacs,longbibliography]{revtex4-2}

\usepackage{tabularx}
\usepackage[latin1]{inputenc}
\usepackage{graphicx}
\usepackage{latexsym}
\usepackage{amssymb}
\usepackage{amsmath}
\usepackage{color}
\usepackage{tikz}
\usepackage{lipsum}
\usepackage{soul}
\usepackage{cancel}

\usepackage{amssymb}
\begin{document}
\def\d{{\rm d}}
\def\ex{{\rm e}}
\def\im{{\rm i}}
\def\e{{\bf e}}
\def\E{{\bf E}}
\def\F{{\bf F}}
\def\M{{\bf M}}
\def\J{{\bf J}}
\def\tr{{\rm tr}}
\def\Ecal{\mathcal{E}}
\def\Ncal{\mathcal{N}}
\def\Lcal{\mathcal{L}}
\def\Jcal{\mathcal{J}}
\def\Bcal{\mathcal{B}}
\def\barBcal{{\bar{\mathcal{B}}}}
\def\Ncal{\mathcal{N}}
\def\barNcal{{\bar{\mathcal{N}}}}
\def\Qcal{\mathcal{Q}}
\def\barQcal{{\bar{\mathcal{Q}}}}
\def\smalze{{\scriptscriptstyle{(0)}}}
\def\smalun{{\scriptscriptstyle{(1)}}}
\def\smaln{{\scriptscriptstyle{(n)}}}
\def\smalf{{\scriptscriptstyle{f}}}
\def\smali{{\scriptscriptstyle{i}}}
\def\smalif{{\scriptscriptstyle{if}}}
\def\beq{\begin{eqnarray}}
\def\eeq{\end{eqnarray}}
%\newcommand{\stc}[1]{\textcolor{magenta}{\st{#1}}}
%\newcommand{\tc}[1]{\textcolor{magenta}{#1}}


\title{Description of a stochastic system by a nonadapted stochastic process}
\author{Piero Olla}
\thanks{Email address for correspondence: olla@dsf.unica.it}
\affiliation{ISAC-CNR and INFN, Sez. Cagliari, I--09042 Monserrato, Italy.}

\begin{abstract}
An approach for the description of stochastic systems is derived.
Some of the variables in the system are studied forward in time, others 
backward in time. The approach 
is based on a perturbation expansion in the strength of the coupling between forward and 
backward variables,
and is well suited for situations in which initial and final conditions
are imposed on different components of the system
and the coupling between those components is weak.
The form of the stochastic equations
in our approach is determined by requiring that they generate the
same statistics obtained in a forward description of the dynamics. 

Numerical tests are carried out on a few simple two-degree-of-freedom systems.
The merit and the difficulties of the
approach are discussed and compared to more traditional strategies based on transition
path sampling and simple shooting algorithms.

\end{abstract}
\maketitle

\section{Introduction}
\label{Introduction}
Stochastic systems are described by equations whose form depend on how the state of the system
is measured, meaning, whether in an experiment, knowledge of the past state of the system or some 
different condition is assumed.  The simplest example is provided by the Langevin dynamics
\beq
\dot E+\Gamma E\d t=\xi,\qquad \langle\xi(t)\xi(0)\rangle=2\Ecal\delta(t),
\label{Langevin}
\eeq
in which the sign of the relaxation coefficient $\Gamma$ depends on whether
we are interested in determining 
the future evolution, or, as in backward induction \cite{puterman},  
the previous history of the system. In the two cases,
the solution of the differential equation can be expressed as an integral over past (future) values 
of the noise $\xi$, corresponding 
to causal (anti-causal) response of the system to external forcing \cite{evans96}.
In mathematical language, one says that the stochastic process is adapted to the forward 
(backward) filtration induced by the Wiener 
process associated with $\xi$ \cite{vanhandel}. The situation differs from the one in 
deterministic (Newtonian) 
mechanics, in which the form of the equations does not depend on the  experiment one
wishes to carry out on the system. 

A forward description allows us in a natural way to take into account initial conditions. 
In the same way, a backward description allows us more easily to take into account final conditions.
It is possible, however, to imagine more general---nonadapted---descriptions of the 
stochastic dynamics.  In fact, there are infinite such descriptions; in the
case of Eq. (\ref{Langevin}), for instance, the same dynamics could be generated by any 
equations in the form,
in frequency space, $G_\omega E_\omega=\xi_\omega$, $\langle\xi_\omega\xi_{\omega'}\rangle=
4\pi\Ecal\delta(\omega+\omega')$, where $|G_\omega|^2= \Gamma^2+\omega^2$.

It has been suggested that nonadapted stochastic processes could be used to describe a stochastic 
dynamics conditioned at multiple times 
\cite{nualart}, and new forms of stochastic calculus have been introduced to deal with 
this type of problems \cite{oksendal}. 
The simplest example of a conditioned stochastic system is the Brownian bridge, 
i.e. a Brownian motion in which the initial and final positions of the particle are supposedly
given. 
The problem was initially studied by Schr\"odinger in the context of a  possible interpretation
of quantum mechanics as a statistical theory with boundary conditions 
in the past and the future \cite{schrodinger31}.  The concept was later extended to more
general systems, typically described by stochastic differential equations (SDEs),
with applications ranging from  mathematical finance \cite{glasserman}, to ecology \cite{horne07}, 
to optimization theory \cite{glanzer20}. Such
extensions of the concept of a Brownian bridge are sometimes referred as a
stochastic bridge \cite{drummond17}.

The mathematical theory for the solution of conditioned SDE's is well established
\cite{pardoux90}.
In practice, however, such problems require a sampling procedure,
which can be numerically demanding if the number of variables 
is large, or if the final condition is a low probability state for the system.
It is still possible, through a Doob transform \cite{doob57,orland11}, 
to write a forward SDE with a 
modified drift that steers trajectories to the imposed final conditions, but the new drift
depends on the same probability of the state of the system conditioned to its final state
that one aims to determine.
To overcome such difficulties, various methods have been devised, either based on 
some form of importance sampling \cite{kloeden,glasserman00,bellet15},
or on the adoption of transition path sampling algorithms \cite{dellago98,crooks01}

In this paper, we focus on a particular version of an ``incomplete'' stochastic bridge, in which
initial and final conditions are imposed separately on different  
subsets of the variable describing the system.
In a large deviation context \cite{chetrite15,derrida19}, e.g.
one may be interested in studying
the development of a large fluctuation in a particular region of a system,
conditioned to the occurrence of some other event, say, the decay of
another large fluctuation elsewhere in the system.
Situations of physical interest include protein folding \cite{mey14}, chemical reactions 
\cite{delarue17,dykman94}, as well as extreme events in the atmosphere \cite{laurie15}.
More specifically, one may wish, e.g. to have an indicator of the occurrence of a future rare 
event in a component of a system (say, a heat wave in a particular region of the planet), 
from the modification of the dynamics in another portion of the system (the atmospheric
circulation in another part of the globe).
The matter has some intriguing aspects, as the interaction of portions of a system where
large fluctuations are developing or decaying can be seen as the interaction between 
portions of the system with opposite (local) arrows of time: positive
where the fluctuation is decaying, negative where the fluctuation is 
developing. Similar issues were studied in a cosmological context in \cite{schulman99}.

Incomplete stochastic bridges provide an example of systems for which a nonadapted description
can cure most difficulties generated by low probability final conditions.
What we want to explore is the possibility of a description, which, in the limit
of vanishing interaction between components of the system with opposite conditioning, becomes
forward for the part of the system conditioned in the past and backward for the part conditioned 
in the future. For finite interaction, the description of the two components will not be purely 
forward or backward, and the system response to external
stimuli will not have well-defined causality properties;
what is lost because of the lack of a physical causal framework, however, is gained witha
low probability final conditions being treated perturbatively around a backward
description of the dynamics, which by construction does not require sampling. We shall
speak in this case, of a mixed backward-forward description of the dynamics (concisely, a mixed 
description), to be opposed to the standard forward description afforded by
Eq. (\ref{Langevin}), and to its time-reversed backward version.

We are going to limit our analysis to 
reversible systems. The simplest example
is that of two bodies exchanging heat with one another and with a thermostat, a 
process that in recent years has attracted the attention of researchers interested 
in fluctuation theorems \cite{jarzynski04,ciliberto13}, and calorimetric experiments 
in mesoscopic systems \cite{averin10,golubev15}.

It is interesting to note that in a mixed description, the two bodies
will see the heat flowing, relative to their own time,
from one to the other with an identical sign, i.e., simultaneously in or out from both.
Such a condition could not be accommodated by
a change of sign of a relaxation coefficient, as in the
shift from Eq. (\ref{Langevin}) to its time-reversed version; a non-trivial redefinition
of the concept of heat exchange is required.

The paper is organized as follows. In Sec. II, the mixed backward-forward approach is introduced
by considering an infinite time horizon. Section III  examines the problem of
boundary conditions. In Sec. IV, the approach is applied to the case of linear dynamics.
In Sec. V, the possible application of the technique to Monte Carlo evaluation of statistical
quantities is discussed, and some applications and tests are presented.
Section VI is devoted to the conclusions.
Technical details are discussed in Appendixes A--C.

\section{Mixed backward-forward approach}
\label{Mixed backward-forward}
We consider the simplest possible example of a system with two degrees of freedom $E_{1,2}$
weakly coupled through correlation in the noise and an interaction component in the 
drift. The dynamics is described in the forward picture by the It\^o SDE
\beq
\dot E_j+F_j(\E)=\xi_j,
\quad
\langle\xi_j(t)\xi_k(0)\rangle=2\Ecal\Xi_{jk}\delta(t),
\label{forward}
\eeq
where
\beq
F_j(\E)=F_j^\smalze(E_j)+gF_j^\smalun(\E),
\label{F}
\eeq
and
\beq
\Xi_{11}=\Xi_{22}=1,
\quad
\Xi_{12}=\Xi_{21}=-g.
\label{Xi}
\eeq

We assume a reversible dynamics, which means that
$X_{jk}F_k\equiv (\Xi^{-1})_{jk}F_k$ must be a gradient \cite{risken}.
This implies
\beq
(X_{1j}\partial_{E_2}-X_{2j}\partial_{E_1})F_j^\smalun=0.
\label{reversible}
\eeq
From Eq. (\ref{Xi}) we find
\beq
X_{11}=X_{22}= \frac{1+g}{1+2g},
\quad
X_{12}=X_{21}=\frac{g}{1+2g},
\label{X}
\eeq
and by combining with Eq. (\ref{reversible}),
\beq 
F_1^\smalun(\E)&=&-F_2^\smalun(\E)
\nonumber
\\
&=&F_1^\smalze(E_1)-F_2^\smalze(E_2)+H(E_1-E_2).
\label{interaction}
\eeq

We seek a mixed backward-forward representation of the dynamics in
Eqs. (\ref{forward}-\ref{interaction}) as a perturbation expansion in $g$, which
we will content ourselves to carry out up to first order. 
The stochastic equations in the mixed picture can be written in the form
\beq
&&\dot E_j+M_j=\xi_j,
\quad
\langle\xi_j(t)\xi_k(0)\rangle=2\Ecal\Xi_{jk}\delta(t),
\nonumber
\\
&&M_j=M_j^\smalze+gM_j^\smalun+\ldots,
\label{mixed}
\eeq
where to lowest order we impose
\beq
M_1^\smalze=F_1^\smalze(E_1),
\quad
M_2^\smalze=-F_2^\smalze(E_2).
\label{M_0}
\eeq
Note that we have assumed identical statistics for the noise in the two pictures, a condition which
is interpreted as a requirement of invariance for
changes of stochastic description at the scale of the noise correlation time.

To determine the interaction terms $M_j^\smalun$, we require identity 
of the statistics in the two representations, 
$\rho_F[\E]=\rho_M[\E]$, where $\rho_F[\E]$ and $\rho_M[\E]$ are 
the functional probability density functions (PDF)
for the trajectories in the two pictures \cite{onsager53,graham77,dykman79}.
Adoption of the It\^o prescription in the forward description guarantees that we do not
have to care about Jacobian factors in the PDF:
\beq
\rho_F[\E]&=&\Ncal\exp\Big[-\frac{1}{2\Ecal}\int\d t\,\Lcal_F\Big],
\nonumber
\\
\Lcal_F&=&\frac{1}{2}X_{jk}(\dot E_j+F_j)(\dot E_k+F_k).
\label{L_F}
\eeq
Since in this work
we are not going to discuss how the system response to external forcing changes
from one description to the other, 
we are not going to adopt a two-field representation for 
$\rho$, such as in the standard Martin-Siggia-Rose approach \cite{martin73}.

To determine the PDF in the mixed picture, we need to select a prescription on the 
stochastic differentials. We interpret Eq. (\ref{mixed}) at the discrete scale as
\beq
\Delta_+E_1(n)+M_1(n)\Delta t=\Delta_+W_1(n),
\label{discrete mixed 1}
\\
\Delta_-E_2(n)-M_2(n)\Delta t=\Delta_-W_2(n),
\label{discrete mixed 2}
\eeq
where $\Delta_\pm$ indicate forward and backward time increments
\beq
\Delta_\pm f(n)=f(n\pm 1)-f(n),
\nonumber
\eeq
and $\Delta_\pm W_{1,2}$ are Wiener increments, 
on which we impose cross correlations at lagged
times:
\beq
&&\langle(\Delta_+ W_1)^2\rangle= \Ecal\Xi_{11}\Delta t,
\quad
\langle(\Delta_- W_2)^2\rangle= \Ecal\Xi_{22}\Delta t,
\nonumber
\\
&&\langle\Delta_+W_1(n)\Delta_-W_2(n+1)\rangle=-\Ecal\Xi_{12}\Delta t.
\label{lagged noise}
\eeq
The adopted prescription reduces, in the decoupled limit $g\to 0$, to an It\^o prescription for 
$E_1$ and a final point prescription for $E_2$, the last one equivalent to an It\^o prescription
in the backward time $t_B=T-t$ ($T$ arbitrary). 
This tells us that for $g\to 0$, the PDF $\rho_M[\E]$ does 
not contain a Jacobian factor. 
For finite $g$, however, $\Delta E_1(n)$ and $\Delta E_2(n)$ receive contributions from terms 
$E_2(n+1)$ and $E_1(n-1)$, anticipating with respect to $t$ in one
case, with respect to $t_B$ in the other. A nontrivial 
Jacobian contribution
$J[\E]=|\det[\partial\Delta_\pm W_j(n)/\partial E_k(m)]|$ 
is therefore expected.
We derive the form of this contribution in 
Appendix \ref{Calculation of}. 
The result of the calculation is  given in Eq. (\ref{we get}), and can be expressed
in continuous time as follows:
\beq
J&=&\exp\Big[-\frac{g}{2\Ecal}\int\d t\,\Lcal_J\Big],
\label{J}
\\
\Lcal_J(t)&=&-2\Ecal\Big[\int_t^{+\infty}\d t'\,
\frac{\delta M^\smalun_1(t')}{\delta E_1(t)}
\nonumber
\\
&-&\int_{-\infty}^t\d t'\,\frac{\delta M^\smalun_2(t')}{\delta E_2(t)}\Big]+O(g).
\label{L_J}
\eeq
The procedure to obtain $\rho_M$ from the PDF for the noise history is identically 
to that in the forward case, and the result is
\beq
\rho_M[\E]&=&\Ncal\exp\Big[-\frac{1}{2\Ecal}\int\d t\,(\Lcal_M+g\Lcal_J)\Big],
\nonumber
\\
\Lcal_M&=&\frac{1}{2}X_{jk}(\dot E_j+M_j)(\dot E_k+M_k),
\label{L_M}
\eeq
where $\dot E_1+M_i\equiv \Delta_+E_1(n)/\Delta t+M_1(n)$ and 
$\dot E_2+M_2\equiv \Delta_+E_2(n)/\Delta t+M_2(n+1)$.

Identity of the statistics in the forward and mixed pictures is established through
\beq
\Lcal_F=\Lcal_M+g\Lcal_J.
\label{identical statistics}
\eeq
Substitution of Eqs. (\ref{L_F}) and (\ref{L_J}-\ref{L_M}) into 
Eq. (\ref{identical statistics}) yields an equation involving time derivatives $\dot E_j$. 
To deal with such terms, we recall the stochastic integration by parts formula
\beq
X_{ij}\frac{\d V_jE_j}{\d t}=X_{jk}[\dot E_kV_k+E_k\dot V_j]+\Ecal\partial_{E_j}V_j,
\label{Ito}
\eeq
which is a consequence of It\^o's lemma \cite{vanhandel}.

Let us consider first the $O(g^0)$ part of Eq. (\ref{identical statistics}). To use
Eq. (\ref{Ito}), we must convert all terms in Eq. (\ref{identical statistics}) 
to a common It\^o prescription. We have from Eqs. (\ref{X}), (\ref{M_0}),
(\ref{L_F}) and (\ref{L_M}):
\beq
\Lcal^\smalze_M-\Lcal^\smalze_F&=&[F_2^\smalze(n)+F_2^\smalze(n+1)]
\frac{\Delta_-E_2(n+1)}{\Delta t}
\nonumber
\\
&=&-2\dot E_2F_2^\smalze(E_2)+\Ecal\partial_{E_2}F_2^\smalze(E_2),
\nonumber
%\label{Zakai}
\eeq
where the second line in the equation is understood in the It\^o sense, and the last term 
is a Zakai-Wong correction \cite{vanhandel}. Exploiting Eq. (\ref{Ito}),
with ${\bf V}=(0,F_2^\smalze)$ and $X_{22}^\smalze=1$, yields the expression
\beq
\Lcal^\smalze_M-\Lcal^\smalze_F=-\frac{\d F_2^\smalze E_2}{\d t},
\label{order zero}
\eeq
which can then be eliminated from Eq. (\ref{identical statistics}),
at the price of a redefinition of the probability of states at $t=\pm\infty$.

We repeat the operations leading to Eq. (\ref{order zero}),
with the $O(g)$ remnant of Eq. (\ref{identical statistics}).
After substitution into Eq. (\ref{identical statistics}), of Eqs.
(\ref{X}-\ref{interaction}), (\ref{L_F}) and (\ref{L_M}), 
and after some straightforward algebra, we reach the result
\beq
&&\Big[E_1(n)\frac{\d}{\d t}-F_1^\smalze\Big]\hat M_1(n)
\nonumber
\\
&+&\Big[E_2(n+1)\frac{\d}{\d t}+F_2^\smalze\Big]\hat M_2(n)
\nonumber
\\
&=&2(F_1^\smalze+H)F_2^\smalze+\Lcal_P+\Lcal_I+\Lcal_J,
\label{order one}
\eeq
where the discrete scale shorthand 
$\d f(n)/\d t \equiv \dot f(n) =\Delta_+f(n)/\Delta t$ continues to be used, we have defined
\beq
\hat M_1(n)&=&M_1^\smalun(n)
\nonumber
\\
&-&F_1^\smalze(n)-F_2^\smalze(n+1)-H(n),
\label{M_1}
\\
\hat M_2(n+1)&=&M_1^\smalun(n+1)
\nonumber
\\
&+&F_1^\smalze(n)+F_2^\smalze(n+1)+H(n),
\label{M_2}
\eeq
together with
\beq
\Lcal_P(n)&=&-2\dot E_1(n)[F_2^\smalze(n)+F_2^\smalze(n+1)]
\nonumber
\\
&+&2\dot E_1(n)[F_2^\smalze(n)-F_2^\smalze(n+1)]
\label{L_P}
\eeq
and
\beq
\Lcal_I(n)=-\Ecal\partial_{E_j}(n)\hat M_j(n),
\label{L_I}
\eeq
and we recall $H=H(E_1-E_2)$ is the part of the interaction explicitly dependent on the 
difference $E_1-E_2$ [see Eq. (\ref{interaction})].
The term $\Lcal_P$ in Eq. (\ref{order one}) 
arises from the discrepancies in the prescriptions in Eqs. 
(\ref{discrete mixed 1}) and (\ref{discrete mixed 2}); the term $\Lcal_I$ in the 
same equation is the It\^o correction arising from application of Eq. (\ref{Ito}) to
the operation of shifting time derivatives in the terms $\dot E_{1,2}\hat M_{1,2}$ in
$\Lcal^\smalun_F-\Lcal_M^\smalun$.
We verify in Appendix \ref{Prescription issues} 
that $\Lcal_P$, $\Lcal_I$ and $\Lcal_J$ can all be disregarded.
Back to continuous time, Eq. (\ref{order one}) becomes then
equivalent to the system of differential equations
\beq
&&\Big(\frac{\d}{\d t}-\frac{F_1^\smalze}{E_1}\Big)\hat M_1
=\frac{F_1^\smalze F_2^\smalze}{E_1}+E_2Q,
\label{ODE1}
\\
&&\Big(\frac{\d}{\d t}+\frac{F_2^\smalze}{E_2}\Big)\hat M_2
=\frac{F_2^\smalze}{E_2}(F_1^\smalze+2H)-E_1Q,
\label{ODE2}
\eeq
where $Q=Q(\E,t)$ is arbitrary.
We can solve Eqs. (\ref{ODE2}) and (\ref{ODE2}) explicitly. By introducing
\beq
R_{1,2}(t)&=&F_{1,2}^\smalze(E_{1,2}(t))/E_{1,2}(t),
\label{R_j}
\\
S_1(t)&=&R_1(t)F_2^\smalze(E_2(t))+E_2(t)Q(t),
\label{S_1}
\\
S_2(t)&=&R_2(t)[F_1^\smalze(E_1(t))
\label{S_2}
\nonumber
\\
&+&2H(E_1(t)-E_2(t))]-E_1(t)Q(t),
\eeq
we can write
\beq
&&\hat M_1(t)=-\int_t^{+\infty}\d\tau\,S_1(t)\exp\Big[-\int_t^\tau\d\tau'\,R_1(\tau')\Big],
\label{hat M_1}
\\
&&\hat M_2(t)=\int_{-\infty}^t\d\tau\,S_2(t)\exp\Big[-\int_\tau^t\d\tau'\,R_2(\tau')\Big],
\label{hat M_2}
\eeq
We see that different choices of the arbitrary function $Q(\E,t)$ allow us to shift the weight of
the nonlocal contribution to the dynamics between $\hat M_1$ and $\hat M_2$.

We summarize the main results of the section.
The stochastic system is described in the mixed picture by a system of 
stochastic equations [Eqs. (\ref{mixed}) and (\ref{M_0})],
involving new variables $\hat M_j$ [Eqs. (\ref{M_1}-\ref{M_2}) and (\ref{ODE1}-\ref{ODE2})]
that are
anticipating relative to the proper time of the respective component: the forward
time $t$ for $j=1$, the backward time $t_B=T-t$ for $j=2$. The new description thus becomes
nonlocal in time, a situation which becomes manifest if the new variables are
expressed as functionals of the old ones through Eqs. (\ref{hat M_1}-\ref{hat M_2}), thus
turning Eq. (\ref{mixed}) into a system of integrodifferential stochastic equations.

The upshot is that a naive perturbative approach based on an iterative solution of
purely forward and backward equations in separate portions of the system,
using at each iteration the values of the variables outside that portion,
obtained in the previous iteration would not work; modified equations must be
used, whose form at $O(g)$ is given in Eqs. (\ref{mixed}), (\ref{M_0}), 
(\ref{M_1}-\ref{M_2}) and (\ref{ODE1}-\ref{ODE2}).

\section{Treatment of boundary conditions}
\label{Treatment of boundary}
We have derived the stochastic equations in the mixed picture, Eqs. (\ref{mixed}), (\ref{M_0}),
(\ref{M_1}-\ref{M_2})
and (\ref{ODE1}-\ref{ODE2}), in an infinite time domain. For the same reason a forward
SDE in the form of Eq. (\ref{Langevin}) cannot directly be utilized to take into account final 
conditions, the equations derived in the previous section cannot be used, in their
current form, for the treatment of boundary condition problems. 

We consider the following boundary conditions
\beq
\Bcal\equiv\Big\{\Bcal_1,\Bcal_2\Big\}=\Big\{E_1(t_i)=E_1^\smali,E_2(t_f)=E_2^\smalf\Big\},
\label{Bcal}
\eeq
which provide us with the simplest example of incomplete stochastic bridge. 

A natural extension of the approach
in Sec. \ref{Mixed backward-forward} would be to solve perturbatively Eq. (\ref{mixed}),
$\E=\E^\smalze+g\E^\smalun+\ldots$, and to enforce Eq. (\ref{Bcal}) order by order in the expansion.
Unfortunately, while it is straightforward to impose Eq. (\ref{Bcal}) on $\E^\smalze$, we
have no control on $\E^\smalun$; in fact,  $E_1^\smalun(t_i)$ and $E_2^\smalun(t_f)$
are determined by the behavior of $\bar\E$ outside $[t_i,t_f]$, and their value is in
general non-zero. The evaluation of statistical quantities must therefore
include correction terms accounting for the fact 
that the boundary conditions in Eq. (\ref{Bcal}) can only be enforced
to $O(g^0)$.

The algebra is somewhat lighter if instead of expanding in $g$, we
decompose $\E$ in contributions from the noise and from the $O(g)$ part of the drift:
\beq
\E\simeq\bar\E+g\tilde\E+O(g^2),
\label{Edecomp}
\eeq
where $\bar\E$ and $\tilde\E$ obey
\beq
\dot{\bar E}_j+M_j^\smalze=\xi_j
\label{Ebar}
\eeq
and
\beq
\dot{\tilde E}_j+\tilde E_k\partial_{\bar E_k}M_j^\smalze+M^\smalun_j=0.
\label{Etilde}
\eeq
Note that since $\bar E_1$ and $\bar E_2$ only depend on the past (future) history of the noise, 
the following condition of statistical independence is going to hold:
\beq
\rho_M(\bar E_1(t)\bar E_2(t'))=\rho_M(\bar E_1)\rho_M(\bar E_2),
\quad
t'>t.
\label{independence}
\eeq
The condition does not hold in the forward picture, therefore $\rho_M[\bar\E]
\ne\rho_F[\bar\E]$. Since all the following calculations are
carried out in the mixed picture, however, for lighter notation, we shall drop subscript 
$M$ on all averages and PDF's involving $\bar\E$.

The statistics of the conditioned problem is contained in the generating functional
\beq
Z[\J]&=&\Big\langle\exp\Big[\im\int\d t\,\J(t)\cdot\E(t)\Big]\Big|\Bcal\Big\rangle
\label{Z}
\nonumber
\\
&=&\Qcal\Big\langle\delta_\Bcal\exp\Big[\im\int\d t\,\J(t)\cdot\E(t)\Big]\Big\rangle,
\eeq
where 
\beq
\Qcal^{-1}=\langle\delta_\Bcal\rangle\equiv\rho(\Bcal),
\label{Ncal}
\eeq
and $\delta_\Bcal$ is the Dirac delta enforcing the boundary condition $\Bcal$. 

Let us indicate
\beq
\barBcal\equiv\Big\{\barBcal_1,\barBcal_2\Big\}=\Big\{\bar E_1(t_i)=
E_1^\smali,\bar E_2(t_f)=E_2^\smalf\Big\},
\label{barBcal}
\eeq
and decompose $Z$ as in Eq. (\ref{Edecomp}),
\beq
Z[\J]=\bar Z[\J]+g\tilde Z[\J]+O(g^2),
\label{Zdecomp}
\eeq
where
\beq
\bar Z[\J]&=&\Big\langle\exp\Big[\im\int\d t\,\J(t)\cdot\E(t)\Big]\Big|\barBcal\Big\rangle
\nonumber
\\
&=&\barQcal\Big\langle\delta_\barBcal\exp\Big[\im\int\d t\,\J(t)\cdot\E(t)\Big]\Big\rangle,
\label{Zbar}
\\
\barQcal^{-1}&=&\langle\delta_\barBcal\rangle\equiv\rho(\barBcal),
\label{barNcal}
\eeq
and
\beq
g\tilde Z[\J]&=&\Big(\frac{\Qcal}{\barQcal}-1\Big)
\Big\langle\exp\Big[\im\int\d t\,\J(t)\cdot\bar\E(t)\Big]\Big|\barBcal\Big\rangle
\nonumber
\\
&+&\barQcal\Big\langle(\delta_\Bcal-\delta_\barBcal)
\exp\Big[\im\int\d t\,\J(t)\cdot\bar\E(t)\Big]\Big\rangle.
\label{Ztilde}
\eeq
Thus, $\bar Z[\J]$ 
contains the statistics that would be obtained if the boundary
conditions were enforced on $\bar\E$ instead of $\E$, and $\tilde Z[\J]$ 
contains the correction.

The correction term $\tilde Z[\J]$ is evaluated in Appendix \ref{Correction terms}. 
Let us introduce the notation 
$\langle \tilde\E\exp(\im\int\J\cdot\E)|\barBcal\rangle^c
\equiv\langle \tilde\E\exp(\im\int\J\cdot\E)|\barBcal\rangle-\langle \tilde\E|\barBcal\rangle
\langle \exp(\im\int\J\cdot\E)|\barBcal\rangle$, and indicate
\beq
&&\nabla_\smalif=(\partial_{E_1^\smali},\partial_{E_2^\smalf}),
\quad
\tilde\E_\smalif=(\tilde E_1(t_i),\tilde E_2(t_f)),
\nonumber
\\
&&\F^\smalze_\smalif=(F_1^\smalze(E_1^\smali),F_2^\smalze(E_2^\smalf)).
\label{smalif}
\eeq
A calculation detailed in Appendix \ref{Correction terms} allows us then to evaluate
the correction term  $\tilde Z[\J]$ as follows
\beq
\tilde Z[\J]&=&\Big[\frac{\F^\smalze_\smalif}{\Ecal}-\nabla_\smalif\Big]
\cdot
\Big\langle\tilde\E_\smalif\exp\Big[\im\int\d t\,\J(t)\cdot\bar\E(t)\Big]\Big|\barBcal\Big\rangle^c
\nonumber
\\
&-&\langle\tilde\E_\smalif|\barBcal\rangle\cdot\nabla_\smalif
\Big\langle\exp\Big[\im\int\d t\,\J(t)\cdot\bar\E(t)\Big]\Big|\barBcal\Big\rangle.
\label{Ztildefin}
\eeq
Substituting Eqs. (\ref{Zbar}) and (\ref{Ztildefin}) into Eq. (\ref{Zdecomp}) produces
the final expression
\beq
Z[\J]&=&[1-g\langle\tilde\E_\smalif|\barBcal\rangle\cdot\nabla_\smalif]
\Big\langle\exp\Big[\im\int\d t\,\J(t)\cdot\E(t)\Big]\Big|\barBcal\Big\rangle
\nonumber
\\
&+&g\Big[\frac{\F^\smalze_\smalif}{\Ecal}-\nabla_\smalif\Big]\cdot
\Big\langle\tilde\E_\smalif\exp\Big[\im\int\d t\,\J(t)\cdot\bar\E(t)\Big]\Big|\barBcal\Big\rangle^c
\nonumber
\\
&+&O(g^2).
\label{Z[J]}
\eeq
%The statistics of the system is thus described in the mixed picture by Eqs. (\ref{mixed}), 
%(\ref{M_0}), (\ref{ODE1}-\ref{ODE2}) and (\ref{Z[J]}).

From Eq. (\ref{Z[J]}) we can obtain expressions for the conditional averages 
$\langle\bar E_{1,2}|\barBcal\rangle$.
We can exploit Eq. (\ref{independence}) to write $\langle\bar E_{1,2}|\barBcal\rangle
=\langle\bar E_{1,2}|\barBcal_{1,2}\rangle$. We introduce the notation for the fluctuations
\beq
\e(t)=\E(t)-\langle\E(t)|\barBcal\rangle,
\nonumber
\eeq
with similar definitions holding for $\bar\e$ and $\tilde\e_\smalif$, and find
\beq
\langle E_1(t)|\Bcal\rangle&=&[1-g\langle\tilde E_1(t_i)|\barBcal\rangle\partial_{E_1^\smali}]
\langle\bar E_1(t)|\barBcal_1\rangle
\nonumber
\\
&+&g\Big[\frac{\F^\smalze_\smalif}{\Ecal}-\nabla_\smalif\Big]\cdot
\langle\tilde\e_\smalif\bar e_1(t)|\barBcal\rangle
\nonumber
\\
&+&g\langle\tilde E_1(t)|\barBcal\rangle+O(g^2),
\label{cond1}
\\
\langle E_2(t)|\Bcal\rangle&=&[1-g\langle\tilde E_2(t_f)|\barBcal\rangle\partial_{E_2^\smalf}]
\langle\bar E_2(t)|\barBcal_2\rangle
\nonumber
\\
&+&g\Big[\frac{\F^\smalze_\smalif}{\Ecal}-\nabla_\smalif\Big]\cdot
\langle\tilde\e_\smalif\bar e_2(t)|\barBcal\rangle
\nonumber
\\
&+&g\langle\tilde E_2(t)|\barBcal\rangle+O(g^2).
\label{cond2}
\eeq

The presence of terms $\propto\Ecal^{-1}\langle\tilde\e_\smalif\bar e_j(t)|\barBcal\rangle$
in Eqs. (\ref{cond1}) and (\ref{cond2}) tells us that fluctuations remain important also in a
large deviation regime $\Ecal\to 0$. It is worth pointing out that these terms produce 
the dominant
contribution to the error in the numerical evaluation of $\langle\E|\Bcal\rangle$. Indeed,
the error in sample averages $\langle\bar\E|\barBcal\rangle_N$
and  $\langle\tilde\E|\barBcal\rangle_N$ scales with $(\Ecal/N)^{1/2}$, 
where $N$ is the sample size, while the one
in $\Ecal^{-1}\langle\tilde\e\bar e_j|\barBcal\rangle_N$ scales with $N^{-1/2}$.

\section{Application to a linear system}
\label{Application}
Consider the following system of SDE's:
\beq
&&\dot E_1+E_1+g(E_1-E_2)=\xi_1,
\label{forlin1}
\\
&&\dot E_2+E_2+g(E_2-E_1)=\xi_2,
\label{forlin2}
\eeq
with noise statistics obeying Eqs. (\ref{forward}) and (\ref{Xi}). The equations,
for $g>0$, could describe in the forward picture 
a pair of identical bodies that exchange heat 
with a thermostat and (weakly) with one another.

The equations describing the system dynamics in the mixed picture are obtained from Eqs. 
(\ref{mixed}), (\ref{M_0}), (\ref{M_1}-\ref{M_2}) 
and (\ref{hat M_1}-\ref{hat M_2}). We note that $H=0$.
In Eqs. (\ref{ODE1}) and (\ref{ODE2}) we set $Q=0$ [the most 
natural choice given the symmetry of Eqs. (\ref{forlin1}-\ref{forlin2})]
and get
\beq
&&\dot E_1+E_1+g(\hat M_1+E_1+E_2)+\ldots=\xi_1,
\label{mixlin1}
\\
&&\dot E_2-E_1+g(\hat M_2-E_1-E_2)+\ldots=\xi_2,
\label{mixlin2}
\eeq
where
\beq
\hat M_1(t)&=&-\int_t^{+\infty}\d\tau\,E_2(\tau)\ex^{t-\tau},
\label{M_1 lin}
\\
\hat M_2(t)&=&\int_{-\infty}^t\d\tau\,E_1(\tau)\ex^{\tau-t}.
\label{M_2 lin}
\eeq
As expected, the heat transfer between the two bodies loses 
in the mixed picture its original character of a relaxation process.

It is possible to verify that the statistics of $\E$ in the two pictures coincide---as requested
by the theory---by
showing that the expression of the correlation functions $\langle E_j(t)E_k(t')\rangle$ obtained
in the two cases is identical. The calculation is straightforward and is not carried out 
here. Instead, we direct
our attention to the dynamics of the conditioned system, and
verify that the forward and the mixed backward-forward approach lead to identical expressions
for the average trajectories $\langle\E(t)|\Bcal\rangle$.

Let us consider first the forward approach. Thanks to linearity,
the average 
trajectories can be expressed as a superposition
$\langle E_j(t)|\Bcal\rangle=\langle E_j(t)|\Bcal_1\rangle+\langle E_j(t)|\Bcal_2\rangle$. 
From Eqs. (\ref{forlin1}-\ref{forlin2}) we then get, for $t\in [t_i,t_f]$,
\beq
\langle E_a(t)|\Bcal\rangle&=&a_+(t-t_i)E_a^\smali+a_-(t_f-t)E_b^\smalf,
\label{exact1}
\\
\langle E_b(t)|\Bcal\rangle&=&a_-(t-t_i)E_a^\smali+a_+(t_f-t)E_b^\smalf,
\label{exact2}
\eeq
where $a_\pm(t)=[\ex^{-t}\pm\ex^{-(1+2g)t}]/2$;
expanding to first order in $g$:
\beq
\langle E_1(t)|\Bcal\rangle&=&E_1^\smali[1-g(t-t_i)]\ex^{t_i-t}+gE_2^\smalf
\Big[(t_f-t)\ex^{t-t_f}
\nonumber
\\
&-&(t_f-t_i)\ex^{2t_i-t-t_f}\Big]+O(g^2),
\label{condavlin1}
\\
\langle E_2(t)|\Bcal\rangle&=&E_2^\smalf[1-g(t_f-t)]\ex^{t-t_f}
+gE_1^\smali\Big[(t-t_i)\ex^{t_i-t}
\nonumber
\\
&-&(t_f-t_i)\ex^{t_i+t-2t_f}\Big]+O(g^2).
\label{condavlin2}
\eeq
The same expressions could be obtained by a rather lengthy calculation
from the Brownian bridge expression for the statistics
$\rho(\E(t)|\Bcal)=\rho(\Bcal_2|\E(t))\rho(\E(t)|\Bcal_1)/\rho(\Bcal_2|\Bcal_1)$
\cite{schrodinger31}.

To obtain the expression for $\langle\E(t)|\Bcal\rangle$ in the mixed picture, we must
evaluate the conditional
averages $\langle\bar\E|\barBcal\rangle$, $\langle\tilde\E|\barBcal\rangle$ 
and the correlation $\langle\tilde\e_\smalif\bar e_j|\barBcal\rangle$.

By setting $F_j^\smalze=E_j$ in  Eqs. (\ref{Ebar}) and exploiting Eq. 
(\ref{independence}), we find
\beq
\langle\bar E_1(t)|\barBcal\rangle=E_1^\smali\ex^{t_i-t},
\quad
\langle\bar E_2(t)|\barBcal\rangle=E_2^\smalf\ex^{t-t_f}.
\label{avbar}
\eeq
By setting $M_1^\smalun=E_1+E_2+\hat M_1$, $M_2^\smalun=-E_1-E_2+\hat M_2$,
with $\hat M_j$ given in Eqs.  (\ref{M_1 lin}-\ref{M_2 lin}), and substituting the result
in (\ref{Etilde}), we find
\beq
\tilde E_1(t)&=&-\int_{-\infty}^t\d\tau\,\ex^{\tau-t}\Big[\bar E_1(\tau)
+\frac{1}{2}\bar E_2(\tau)\Big]
\nonumber
\\
&+&\frac{1}{2}\int_t^{+\infty}\d\tau\,\ex^{t-\tau}\bar E_2(\tau)
\label{tildelin1}
\eeq
and
\beq
\tilde E_2(t)&=&-\int_t^{+\infty}\d\tau\,\ex^{t-\tau}\Big[\bar E_2(\tau)
+\frac{1}{2}\bar E_1(\tau)\Big]
\nonumber
\\
&+&\frac{1}{2}\int_{-\infty}^t\d\tau\,\ex^{\tau-t}\bar E_1(\tau).
\label{tildelin2}
\eeq
Substituting Eq. (\ref{avbar}) into Eqs. (\ref{tildelin1}-\ref{tildelin2})
we then obtain
\beq
\langle\tilde E_1(t)|\barBcal\rangle&=&-E_1^\smali\Big(\frac{1}{2}+t-t_i\Big)\ex^{t_i-t}
\nonumber
\\
&+&\frac{E_2^\smalf}{2}(t_f-t)\ex^{t-t_f},
\label{avtilde1}
\\
\langle\tilde E_2(t)|\barBcal\rangle&=&-E_2^\smalf\Big(\frac{1}{2}+t_f-t\Big)\ex^{t-t_f}
\nonumber
\\
&+&\frac{E_1^\smali}{2}(t-t_i)\ex^{t_i-t}.
\label{avtilde2}
\eeq

The fluctuation contribution  $\langle\tilde\e_\smalif\bar e_j|\barBcal\rangle$ is proportional,
through Eqs. (\ref{tildelin1}-\ref{tildelin2}), to
$\langle\bar e_i(t)\bar e_j(t')|\barBcal\rangle$.
Thus, since $\langle\tilde\e_\smalif\bar e_j|\barBcal\rangle$ enters Eqs. (\ref{cond1}-\ref{cond2})
already at $O(g)$, 
we need to evaluate  $\langle\bar e_i(t)\bar e_j(t')|\barBcal\rangle$ only
to $O(g^0)$; it is thus irrelevant whether we work in the forward or in the 
backward picture. We immediately find
$\langle\bar e_1\bar e_1|\barBcal\rangle^\smalze=
\langle\bar e_1\bar e_1|\barBcal_1\rangle^\smalze$,
$\langle\bar e_2\bar e_2|\barBcal\rangle^\smalze=
\langle\bar e_2\bar e_2|\barBcal_2\rangle^\smalze$,
and $\langle\bar e_1\bar e_2|\barBcal\rangle=O(g)$; 
the last term can therefore be disregarded. 
Since $e_1(t)$ and $e_1(t')$ are uncorrelated at opposite
sides of $t_i$, and $e_2(t)$ and $e_2(t')$ are uncorrelated at opposite
sides of $t_f$, the only contribution to $\langle\E(t)|\Bcal\rangle$ comes from
\beq
\langle\bar e_1(t)\bar e_1(t')|\barBcal\rangle^\smalze=\frac{\Ecal}{2}\Big[\ex^{-|t-t'|}
-\ex^{-|t+t'-2t_i|}\Big],
\eeq
with $t,t'>t_i$ or $t,t'<t_i$,
and from
\beq
\langle\bar e_2(t)\bar e_2(t')|\barBcal\rangle^\smalze=\frac{\Ecal}{2}\Big[\ex^{-|t-t'|}
-\ex^{-|t+t'-2t_f|}\Big],
\eeq
with $t,t'>t_f$ or $t,t'<t_f$.
By exploiting Eqs. (\ref{tildelin1}-\ref{tildelin2})
we then get the result, for $t\in [t_i,t_f]$,
\beq
\langle\tilde e_1(t_i)\bar e_2(t)|\barBcal\rangle
&=&\frac{\Ecal}{2}\Big[(t-t_i)\ex^{t_i-t}
\nonumber
\\
&-&(t_f-t_i)\ex^{t_i+t-2t_f}\Big]+O(g),
\label{e1e2}
\\
\langle\tilde e_2(t_f)\bar e_1(t)|\barBcal\rangle
&=&\frac{\Ecal}{2}\Big[(t_f-t)\ex^{t-t_f}
\nonumber
\\
&-&(t_f-t_i)\ex^{2t_i-t-t_f}\Big]+O(g),
\label{e2e1}
\\
\langle\tilde e_1(t_i)\bar e_1(t)|\barBcal\rangle&\sim&
\langle\tilde e_2(t_f)\bar e_2(t)|\barBcal\rangle=O(g).
\label{eiei}
\eeq
We can now substitute Eqs. (\ref{avbar}), 
(\ref{avtilde1}-\ref{avtilde2}) and (\ref{e1e2}-\ref{e2e1})
into Eqs. (\ref{cond1}-\ref{cond2}), and verify with little algebra that the same expression 
for $\langle\E|\Bcal\rangle$ in
Eqs. (\ref{condavlin1}-\ref{condavlin2}) is reproduced.

\section{Numerical tests}
\label{Numerical tests}
The numerical solution of the stochastic equations in the mixed picture
has important peculiarities.
Firstly, since different sets of 
variables are integrated into opposite time directions, the evaluation of the
interaction terms require that the histories of the variables $\bar E_j$ 
be already calculated and stored in memory.
Secondly, because of the intrinsic time non-local nature of 
the mixed picture, and the fact that boundary conditions in an incomplete stochastic
bridge are imposed only on one part of the variables,
the stochastic equations for a problem conditioned at $t_i$ and $t_f$, $t_i<t_f$,  
must be solved in a wider domain $[T_i,T_f]\supset [t_i,t_f]$.

The determination of the trajectories, to be used for 
the Monte Carlo evaluation of statistical quantities in Eq. (\ref{Z[J]}), 
therefore, is going to proceed through the following sequence of steps:

\begin{itemize}
\item
Generate 
a noise history $\boldsymbol{\xi}$ in $[T_i,T_f]$, and store it in memory.
\item
Obtain from $\boldsymbol{\xi}_i$ and store in memory the history of $\bar\E$ in the
interval $[T_i,T_f]$
(at this point the history of $\boldsymbol{\xi}$ is not needed anymore).
Since the whole history of $\bar\E$ in the whole interval $[t_i,t_f]$ is required,
forward variables are going to become backward variables in $[T_i,t_i]$ 
and backward variables become forward variables     
in $[t_f,T_f]$. For the
problem considered in Sec. \ref{Treatment of boundary}, this means
that $\bar E_1$ and $\bar E_2$ are going to obey equations
$\dot{\bar E}_1-F_1^\smalze=\xi_1$ and $\dot{\bar E}_2+F_2^\smalze=\xi_2$
in domains $[T_i,t_i]$ and $[t_f,T_f]$, respectively
\item
Again in $[T_i,T_f]$, obtain $\hat\M$ from $\bar\E$. For the system considered in Sec. 
\ref{Mixed backward-forward}, this is accomplished by integrating Eq. (\ref{ODE1}) backward
and Eq. (\ref{ODE2}) forward in time. The domain $[T_i,T_f]$ must therefore be large enough
to guarantee that the effect of the boundary conditions at $T_{i,f}$ on the profile of 
$\hat\M$ in $[t_i,t_f]$ be negligible.
Since each $\tilde E_j$ is is obtained by integrating in the time direction opposite to
that of the corresponding $\hat M_j$, also $\hat\M$ must be stored in memory.
\item
From $\bar\E$ and $\hat\M$, finally obtain also $\tilde\E$; this requires 
integrating the equations for the forward variables from $T_i$ to $t_f$ and those for
the backward variables from $T_f$ to $t_i$. Once more, the domain $[T_i,T_f]$ must
be chosen large enough for the boundary conditions on $\tilde\E$ at $T_{i,f}$ 
not to affect the behavior of the variables in the domain of interest $[t_i,t_f]$.
\end{itemize}

\begin{figure}
\begin{center}
\includegraphics[draft=false,width=8cm]{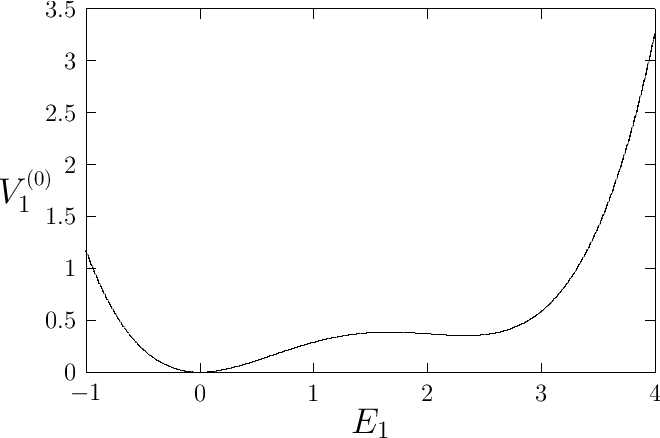}
\caption
{
Profile of the potential $V_1^\smalze$ for system $c$.
}
\label{adfig0}
\end{center}
\end{figure}

We continue to focus our attention on simple two-degree-of-freedom systems, 
and verify that the forward and
the mixed picture generates identical statistics. We carry out our tests
on the deviation of the average trajectories from the decoupled case:
\beq
\langle\E(t)|\Bcal\rangle^\prime=\langle\E(t)|\Bcal\rangle-
\langle\E(t)|\Bcal\rangle^\smalze. 
\eeq

We consider three examples of stochastic system; the following profiles for
the unperturbed drift are adopted:
\beq
F_1^\smalze=E_1,\quad &{\rm system\ }a,
\\
F_1^\smalze=\frac{1}{6}(E_1^3+E_1),\quad &{\rm system\ }b,
\\
F_1^\smalze=\frac{1}{3}\Big(E_1^3-4E_1^2+\frac{35}{9}E_1\Big),\quad &{\rm system\ } c,
\eeq
with $F_2^\smalze=E_2$ in the three cases.
The statistics in Eq. (\ref{forward}) is utilized
for the noise and the condition $H=0$ is imposed in the interaction term.
The values of the remaining parameters are listed
in Table \ref{table1}. 
\begin{table}[h]
\footnotesize
\caption{\label{table1}
Simulation parameters.
}
\begin{tabularx}{\columnwidth}{XX}
\hline
\hline
$t_i=0,$
& $t_f=3$, (system $a$)
\\
$T_i=-17,$
& $T_f=20,$
\\
$E_1^\smali=4$ 
& $E_2^\smalf=1$, (system $a$)
\\
$\Ecal=0.1$,
& $g=0.1$,
\\
$N=10^5$ (sample size).
& 
\\
\hline
\hline
\end{tabularx}
\end{table}

The potential $V_1^\smalze$, $F_1^\smalze=V_1^{\smalze\prime}$, has
for systems $a$ and $b$ the form a single quadratic  
(quartic) well; for system $c$, it is a rather shallow
double potential well, with profile shown in Fig. \ref{adfig0}.

\begin{figure}
\begin{center}
\includegraphics[draft=false,width=8cm]{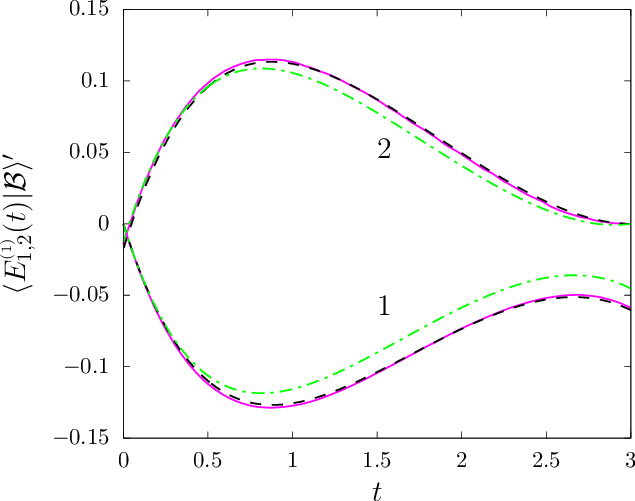}
\caption
{
Mean trajectory deviation from the decoupled limit in the case of system $a$. 
Result of the numerical integration continuous line (magenta online); 
exact profiles from Eqs. (\ref{exact1}-\ref{exact2}) dash-dotted line (green online);
analytical $O(g)$ profiles from
Eqs. (\ref{condavlin1}-\ref{condavlin2}) black dashed line.
}
\label{adfig1}
\end{center}
\end{figure}

In the case of system $a$, analytical expressions for the mean profiles 
of a generic incomplete stochastic bridge are available
[Eqs. (\ref{condavlin1}-\ref{condavlin2})]. As
can be checked in Fig. \ref{adfig1}, the result of numerical simulations basically overlap with
the analytical expressions for the $O(g)$ profiles in Eq. (\ref{condavlin1}-\ref{condavlin2}).

In the case of systems $b$ and $c$, we consider an initial condition problem, with
$\Bcal_1=\{E_1(t_i)=E_1^\smali\}$, and $E_2(t_i)$ distributed according to 
$\rho(E_2(t_i)|\Bcal_1)$. This allows direct comparison of numerical simulations in 
the forward and mixed picture, without having to resort, in the forward 
case, to sampling.

In order to compare the result of simulations in the two pictures,
the time separation $t_f-t_i$ must be sufficiently large to be able to
approximate $\rho(E_2(t_i)|\Bcal)\simeq \rho(E_2(t_i)|\Bcal_1)$. In alternative,
$E_2(t_f)$ must be extracted from the PDF $\rho(E_2(t_f)|\Bcal)$, which, to the 
order in $g$ considered, can be approximated with 
$\rho^\smalze(E_2(t_f))\sim\exp[-(E_2(t_i))^2/(2\Ecal)]$ anyway.	
We note that for the dynamics considered the relation 
$\rho(E_2(t_i))=\rho^\smalze(E_2(t_i))\sim\exp[-(E_2(t_i))^2/(2\Ecal)]$
holds exactly.

\begin{figure}
\begin{center}
\includegraphics[draft=false,width=8cm]{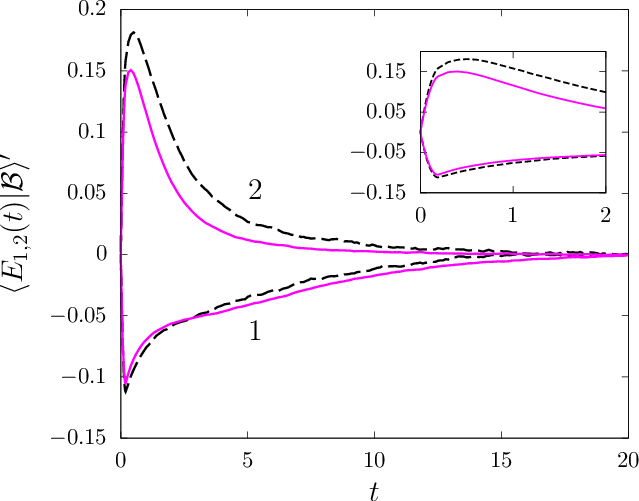}
\caption
{
Mean trajectory deviation from the decoupled limit in the case of system $b$.
Mixed picture black dashed line; forward picture light continuous line (magenta online).
}
\label{adfig3}
\end{center}
\end{figure}
In the case of system $b$, as shown in Fig. \ref{adfig3}, 
the gap between the mean trajectories in the mixed and forward
picture is in the range expected for the values of the coupling $g$ considered.

\begin{figure}
\begin{center}
\includegraphics[draft=false,width=8cm]{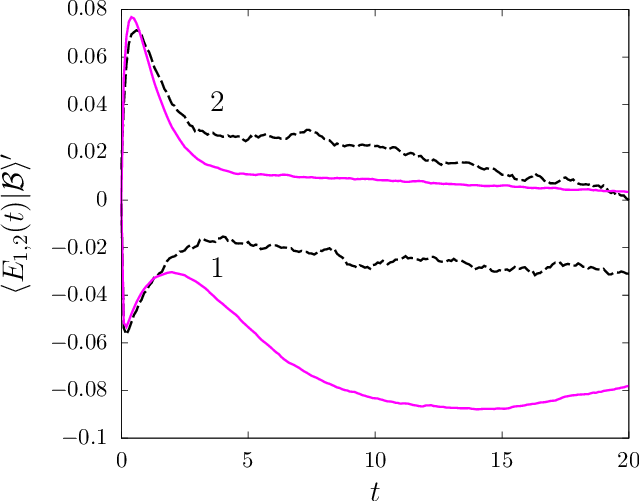}
\caption
{
Same as Fig. \ref{adfig3} in the case of system $c$. 
}
\label{adfig4}
\end{center}
\end{figure}

\begin{figure}
\begin{center}
\includegraphics[draft=false,width=8cm]{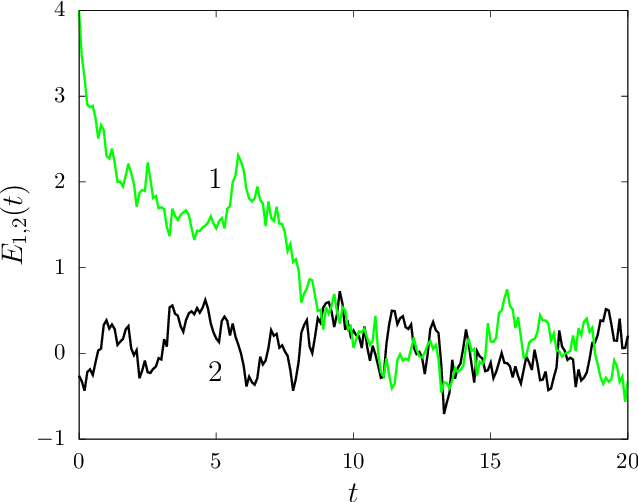}
\caption
{
Individual trajectories for system $c$, obtained from integrating in the forward approach
($E_1$ green online). Notice the long time spent by the system at $E_1\gtrsim 2$, 
before making transition to the potential well at $E_1\approx 0$.
}
\label{adfig5}
\end{center}
\end{figure}

The performance of the mixed approach in the case of system $c$ is much worse.
As shown in Fig. \ref{adfig4}, the mixed approach heavily underestimates 
$\langle E_1(t)|\Bcal_1\rangle^\prime$, which is the contribution 
to the relaxation of $E_1$ to the bottom of the potential well 
at $E_1=0$ from interaction with  $E_2$.
The poor performance of the mixed approach appears to be a
a manifestation of the breakdown of the perturbative ansatz
$gF_2^\smalun\ll F_2^\smalze$ near the potential barrier  at $E_2\approx 1.8$
(see Fig. \ref{adfig0}), where
the contribution from $F_1^\smalun$ to the escape of $E_1$ from the shallow well
to the right is dominant.
The situation is similar to the breakdown of the WKB expansion in the vicinity
of turning points \cite{landau}.
In the present case, the poor performance of the mixed backward-forward approach 
is associated with the long permanence of $E_1$ in the shallow potential well to the
right (see Fig. \ref{adfig5}), 
where the dynamics are dominated by the noise and by the interaction with $E_2$. 

\section{Conclusion}
We have derived a mixed backward-forward approach for the treatment of stochastic
systems, with initial and final conditions imposed on different subsets of the variables 
that describe the dynamics. The main results of the paper are contained in
Eqs. (\ref{ODE1}-\ref{ODE2}) and (\ref{Z[J]}), which provide the form of the SDE's,
together with the procedure for the calculation of conditional averages in our approach.
The interaction between forward and backward variables
is taken into account by the introduction of additional internal degrees of freedom, which
make the dynamics intrinsically nonlocal in time.
%and modify how average quantities are evaluated. 

Nonlocality in time turns the boundary conditions on the system into statistical constraints
in an infinite time domain; this however is precisely what happens, independently of the 
description, in an incomplete stochastic bridge: by construction, 
variables in an incomplete stochastic bridge that do not satisfy boundary conditions
at an end of the bridge, bring information from the outside into the bridge. 
It is thus not too surprising that the most natural description of a system in which
initial and final conditions are imposed on different sets of variables, is nonlocal
in time. 

As regards the cosmological models in \cite{schulman99}, the implication
is that the hypothetical presence of regions in the universe with an arrow of time opposite to
ours would require some form of time nonlocality 
in the interaction between us and ``them''.

We have derived the mixed backward-forward approach in the case of reversible dynamics described
in the forward picture by an SDE with additive noise. 
No additional hypotheses have been made
on the form of the SDE's, except the smallness of the interaction between forward and backward
variables. For the sake of clarity the derivation has been carried out in the simplest possible 
case of a two-variable system; the  generalization to systems with  a higher number of variables,
however, is straightforward. A question that remains open is
the possible extension of the perturbative expansion to higher orders
in the coupling strength. It likewise remains open the question of a possible extension of
the approach to the case of irreversible dynamics.

From the point of view of the Monte Carlo evaluation of statistical quantities, the main
advantage of the approach is the possibility of treating 
final conditions as if they were initial 
conditions in a forward approach. Therefore, as opposed to shooting algorithms,
the approach does not need large samples to treat low probability final states;
at the same time, it does not have the problems
of slow convergence and  difficult handling of final points of
transition path sampling algorithms  \cite{ceperley95,crooks01}.
The price is the increased memory requirement
implicit---as discussed in Sec. \ref{Numerical tests}---in 
a forward-backward description of the dynamics, 
and the constraint of weak interaction between
components of the system with opposite conditioning.

Numerical tests in the simplest possible case of a system with two degrees of
freedom show that the mixed backward-forward approach works as long as there is no crossover 
in the strength of the interaction relative to the unperturbed dynamics. A situation of 
physical interest where the condition is violated is the crossing of a potential barrier.
The situation is similar to that of the WKB method near a turning point, and  the procedure to 
tackle the problem is expected to be the same, namely, to shift to an alternative 
description in the vicinity of the barrier.  How to carry out the procedure, however, remains
unclear at the moment, and the matter deserves further study.

\vskip 10pt
\noindent
{\bf Acknowledgements}
\\
Part of this research was carried out at the Mathematics Department of the University
of Helsinki. The author thanks Prof. Paolo Muratore Ginanneschi for hospitality.
This research was supported in part by the AtMath Collaboration at the University of Helsinki 
and by FP7 EU project ICE-ARC (Grant agreement No. 603887).

\appendix
\section{Calculation of the Jacobian determinant}
\label{Calculation of}
Consider initially a finite time domain $[n_i,n_f]$. We can
dispose the entries of the four-index matrix 
$\partial\Delta_\pm W_j(n)/\partial E_k(m)$ along the rows and
columns of a staggered $2(n_f-n_i-1)\times 2(n_f-n_i-1)$ 
matrix $\Jcal_{\alpha(j,n),\beta(k,m)}$, where indices
$\alpha=1,2,\ldots$ map to
\beq
(j,n)&=&(1,n_i),(2,n_f),(1,n_i+1),(2,n_f-1),
\nonumber
\\
&&\ldots,(1,n_f-1),(2,n_i+1);
\eeq
and indices $\beta=1,2,\ldots$ map to
\beq
(k,m)&=&(1,n_i+1),(2,n_f-1),(1,n_i+2),(2,n_f-2),
\nonumber
\\
&&\ldots,(1,n_f),(2,n_i).
\eeq
Let us decompose 
\beq
\Jcal=\bar\Jcal+g\tilde\Jcal=\bar\Jcal[1+g\hat\Jcal].
\label{decompose}
\eeq
We see from Eq. (\ref{mixed}) that to lowest order in $\Delta t$ $\bar\Jcal$ is a banded matrix 
\beq
\bar\Jcal_{\alpha\beta}=\delta_{\alpha\beta}-\delta_{\alpha,\beta+2}+O(\Delta t),
\eeq
with determinant 
\beq
\det\bar\Jcal=1+O(\Delta t),
\label{det1}
\eeq
and inverse, the lower triangular matrix
\beq
\bar\Jcal^{-1}=
\begin{bmatrix}
1 & 0 &   &   &   &   &        &
\\
0 & 1 & 0 &   &   &   &        &
\\
1 & 0 & 1 & 0 &   &   & \ldots &
\\
0 & 1 & 0 & 1 & 0 &   &        &
\\
1 & 0 & 1 & 0 & 1 & 0 &        &
\\
  &   &   &\vdots &&  &        &
\\
\end{bmatrix}
+O(\Delta t),
\label{triangular}
\eeq
Following standard practice \cite{zinnjustin}, 
we write the Jacobian $J=|\det\Jcal|$ as a Taylor series:
\beq
J&=&|\det\bar\Jcal||\det(1+g\hat\Jcal)|=\exp[\tr\ln(1+g\hat\Jcal)]
\nonumber
\\
&=&\exp(1+g\tr\hat\Jcal)+O(g^2),
\label{standard practice}
\eeq
where we have exploited Eq. (\ref{det1}), and we have disregarded $O(\Delta t)$ terms, 
that vanish in the continuous limit.
Substituting Eqs. (\ref{decompose}) and
(\ref{triangular}) into Eq. (\ref{standard practice}), we get
\beq
\tr\hat\Jcal
&=&\sum_{\alpha\ge 1}\sum_{\beta\ge\alpha}
[\tilde\Jcal_{2\alpha-1,2\beta-1}+\tilde\Jcal_{2\alpha,2\beta}]
\nonumber
\\
&=&\sum_m\Big[
\sum_{n>m}\frac{\partial\Delta W_1(t_n)}{\partial E_1(t_m)}
+\sum_{n<m}\frac{\partial\Delta W_2(t_n)}{\delta E_2(t_m)}
\Big]
\nonumber
\\
&=&\Delta t\sum_m\Big[
\sum_{n>m}\frac{\partial M^\smalun_1(n)}{\partial E_1(m)}
-\sum_{n<m}\frac{\partial M^\smalun_2(n)}{\partial E_1(m)}
\Big]
\nonumber
\\
&+&O(g).
\label{we get}
\eeq
Note that there is no contribution to $\tr\hat\Jcal$ from $M^\smalze_j$, which is consequence
of the fact that there is no contribution in the sums in Eq. (\ref{we get}) from $n=m$. This in
turn is consequence of the fact that $M^\smalze_1(n)$ does not depend on $E_1(n+1)$, and that
$M^\smalze_2$ does not depend on and $E_2(n-1)$. 
Taking in Eq. (\ref{we get}) the two limits $\Delta t\to 0$ and $[t_i,t_f]\to[-\infty,\infty]$ 
finally yields Eqs. (\ref{J}) and (\ref{L_J}).

\section{Prescription issues in the mixed picture}
\label{Prescription issues}
Let us analyze separately the contributions to Eq. (\ref{order one}) from $\Lcal_P$, $\Lcal_I$
and $\Lcal_J$. 
We promptly verify that the first term on the right hand side (RHS) of Eq. (\ref{L_P}) is 
a Stratonovich differential \cite{risken}, 
which brings no It\^o correction when integrating by parts, and that the second term 
is proportional to the noise cross correlation, which is $O(g)$.  Hence, to the order considered,
we can set $\Lcal_P=0$. 

It is difficult to prove in general that $\Lcal_I=\Lcal_J=0$, but
we can verify {\it a posteriori} that the condition is satisfied in the case of
Eqs. (\ref{hat M_1}-\ref{hat M_2}).

In the case of $\Lcal_I$, 
the fact that the argument $\E$ in the functionals $\hat M_j[\E,t]$,
as illustrated in Eqs. (\ref{hat M_1}-\ref{hat M_2}), depends
on time only through a dummy integration variable, guarantees that $\dot{\hat M}_j[\E,t]
=\partial_t\hat M_j[\E,t]$. Hence, substituting $V_j=\hat M_j$ in Eq. (\ref{Ito})
does not generate an It\^o correction, and therefore
$\Lcal_I=0$. 

As regards $\Lcal_J$, Eqs. (\ref{hat M_1}-\ref{hat M_2}) tell us that
$\hat M_1(\tau)$ and $\hat M_2(\tau)$ receive contribution from 
$\E(t>\tau)$ in one case, from $\E(t<\tau)$ in the other. In addition to this, we have seen
by working in discrete time, that $M_j^\smalun(n)$ depends as a function on 
$E_j(m)$ only for $m=n$ [see Eqs. (\ref{M_1}) and (\ref{M_2})].
However, from Eqs. (\ref{L_J}) and (\ref{we get}), $\Lcal\ne 0$
only if $M_1^\smalun(n)$ depends on $E_1(m<n)$, or $M_2^\smalun(n)$ depends on $E_2(m>n)$, 
or both.
Hence, $\Lcal_J=0$.

\section{Correction terms in the evaluation of conditional averages}
\label{Correction terms}
We evaluate the averages in Eqs. (\ref{Zbar}-\ref{Ztilde}) in terms of the PDF
\beq
\rho[\bar\E]=\rho(\bar E_1(t_i),\bar E_2(t_f))
\rho[\bar\E|\bar E_1(t_i),\bar E_2(t_f)].
\label{rho[Ebar]}
\eeq
We start by evaluating the PDF $\rho(\Bcal)$ in Eq. (\ref{Ncal}). We Taylor expand
$\Bcal$ around $\bar\E$ and substitute into $\langle\delta_\Bcal\rangle=\int[\d\bar E]\,
\rho[\bar\E]\, \delta_\Bcal$. Equation (\ref{rho[Ebar]}) gives us then
\beq
\rho(\Bcal)&=&\rho(\barBcal)\Big\{1-g\Big[\nabla_\smalif\cdot\langle\tilde\E_\smalif|\barBcal\rangle
\nonumber
\\
&+&\langle\tilde\E_\smalif|\barBcal\rangle\cdot\nabla_\smalif\ln\rho(\barBcal)\Big]\Big\}
+O(g^2),
\label{rho(Bcal)}
\eeq
where 
\beq
\nabla_\smalif=(\partial_{E_1^\smali},\partial_{E_2^\smalf}),
\qquad
\tilde\E_\smalif=(\tilde E_1(t_i),\tilde E_2(t_f)).
\label{smalif1}
\eeq
In Eq. (\ref{rho(Bcal)}) we can approximate $\rho(\barBcal)\simeq
\rho^\smalze(\barBcal)=\rho^\smalze(\barBcal_1)\rho^\smalze(\barBcal_2)$, where, 
thanks to reversibility of the dynamics,
\beq
\rho^\smalze(\barBcal_1)\sim
\exp\Big\{-\frac{1}{\Ecal}\int^{E_1^\smali}\d\bar E_1\,F_1^\smalze(\bar E_1)\Big\}
\label{rho(barBcal_1)}
\eeq
and
\beq
\rho^\smalze(\barBcal_2)\sim
\exp\Big\{-\frac{1}{\Ecal}\int^{E_2^\smalf}\d\bar E_2\,F_2^\smalze(\bar E_2)\Big\}.
\label{rho(barBcal_2)}
\eeq
By putting together Eqs. (\ref{Ncal}) and (\ref{barNcal}) with Eqs. (\ref{rho(Bcal)}) and
(\ref{rho(barBcal_1)}-\ref{rho(barBcal_2)}), we then get
\beq
\frac{\Ncal}{\barNcal}-1=g\Big[\nabla_\smalif
-\frac{\F^\smalze_\smalif}{\Ecal}\Big]\cdot
\langle\tilde\E_\smalif|\barBcal\rangle+O(g^2),
\eeq
where
\beq
\F^\smalze_\smalif=(F_1^\smalze(E_1^\smali),F_2^\smalze(E_2^\smalf)).
\label{smalif2}
\eeq
We can repeat the procedure with the second line of Eq. (\ref{Ztilde}), to 
obtain 
\beq
\tilde Z[\J]&=&\Big[\frac{\F^\smalze_\smalif}{\Ecal}-\nabla_\smalif\Big]
\cdot
\Big\langle\tilde\E_\smalif\exp\Big[\im\int\d t\,\J(t)\cdot\bar\E(t)\Big]\Big|\barBcal\Big\rangle
\nonumber
\\
&-&\Big\langle\exp\Big[\im\int\d t\,\J(t)\cdot\bar\E(t)\Big]\Big|\barBcal\Big\rangle
\nonumber
\\
&\times& \Big[\frac{\F^\smalze_\smalif}{\Ecal}-\nabla_\smalif\Big]
\cdot\langle\tilde\E_\smalif|\barBcal\rangle,
\eeq
and from here we get Eq. (\ref{Ztildefin}).

\bibliography{sample}

\end{document}